\documentclass[final,5p,times,twocolumn]{elsarticle}

\usepackage{amssymb}
\usepackage{hyperref}
\hypersetup{colorlinks=true}
\usepackage{amsmath}
\usepackage{amsthm}
\usepackage{graphicx}
\usepackage{caption}
\usepackage{subcaption}
\journal{Fusion Engineering and Design}

\begin{document}

\begin{frontmatter}



\title{Nucleation, growth and transport modelling of helium bubbles under nuclear irradiation in lead-lithium with the Self-consistent nucleation theory and surface tension corrections}


\author[1]{J. Fradera\corref{cor1}\fnref{labelc1}}
\author[1]{S.Cuesta-L\'{o}pez\fnref{labelc2}}
\address[1]{Advanced Materials, Nuclear Technology, Applied Nanotechnology, University of Burgos (UBU), Science and Technology Park, I+D+I Building, Room 63, Plaza Misael Bañuelos, s/n, 09001, Burgos, Spain}
\cortext[cor1]{corresponding author}
\fntext[labelc1]{contact: jfradera@ubu.es}
\fntext[labelc2]{correspondence may also be sent to: scuesta@ubu.es}
\newcommand{\LiPb}{Pb15.7Li }
\newcommand{\OF}{OpenFOAM\textsuperscript{\textregistered} }

\begin{abstract}
Helium (He) nucleation in liquid metal breeding blankets of a DT fusion reactor may have a significant impact regarding system design, safety and operation. Large He production rates are expected due to tritium (T) fuel self-sufficiency requirement, as both, He and T, are produced at the same rate. Low He solubility, local high concentrations, radiation damage and fluid discontinuities, among other phenomena, may yield the necessary conditions for He nucleation. Hence, He nucleation may have a significant impact on T inventory and may lower the T breeding ratio.

A model based on the self--consistent nucleation theory (SCT) with a surface tension curvature correction model has been implemented in \OF CFD code. A modification through a single parameter of the necessary nucleation condition is proposed in order to take into account all the nucleation triggering phenomena, specially radiation induced nucleation. Moreover, the kinetic growth model has been adapted so as to allow for the transition from a critical cluster to a macroscopic bubble with a diffusion growth process.

Limitations and capabilities of the models are shown by means of zero-dimensional simulations and sensitivity analyses to key parameters under HCLL breeding unit conditions. Results provide a good qualitative insight into the helium nucleation phenomenon in LM systems for fusion technology and reinforces the idea that nucleation may not be a remote phenomenon, may have a large impact on the system's design and reveals the necessity to conduct experiments on He cavitation.

\end{abstract}


\end{frontmatter}

\newcommand{\LiPb}{Pb15.7Li }
\newcommand{\OF}{OpenFOAM\textsuperscript{\textregistered} }

\section*{Glossary}
\newcommand{\Item}[2]{\item[\textbf{#1\hfill}] #2}
\newcommand{\Itema}[2]{\item[\hfill #1] #2}

\subsection*{Abbreviations}
\begin{list}{}{%
\settowidth{\labelwidth}{\textbf{HCLL}}%
\setlength{\labelsep}{2. em}%
\setlength{\leftmargin}{\labelwidth}%
\addtolength{\leftmargin}{\labelsep}%
\setlength{\rightmargin}{0. cm}%
\setlength{\parsep}{\parskip}%
\setlength{\itemsep}{0. cm}\setlength{\topsep}{0. cm}%
\setlength{\partopsep}{0. cm}}

\Item{CFD}{Computational Fluid Dynamics}
\Item{CNT}{Classical Nucleation Theory}
\Item{DT}{deuterium--tritium}
\Item{EoS}{Equation of State}
\Item{HCLL}{Helium Cooled Lithium-Lead}
\Item{HEN}{heterogeneous nucleation}
\Item{HON}{homogeneous nucleation}
\Item{LM}{liquid metal}
\Item{SM}{structural material}
\Item{T}{tritium}
\end{list}

\subsection*{Greek characters}
\begin{list}{}{%
\settowidth{\labelwidth}{\textbf{HCLL}}%
\setlength{\labelsep}{2. em}%
\setlength{\leftmargin}{\labelwidth}%
\addtolength{\leftmargin}{\labelsep}%
\setlength{\rightmargin}{0. cm}%
\setlength{\parsep}{\parskip}%
\setlength{\itemsep}{0. cm}\setlength{\topsep}{0. cm}%
\setlength{\partopsep}{0. cm}}

\Item{$\alpha$}{void fraction}
\Item{$\theta$}{contact angle}
\Item{$\pi$}{number pi}
\Item{$\rho$}{density}
\Item{$\sigma$}{surface tension}
\Item{$\upsilon_0$}{volume of one atom or molecule}
\Item{$\psi$}{supersaturation ratio}
\end{list}

\subsection*{Latin characters}
\begin{list}{}{%
\settowidth{\labelwidth}{\textbf{HCLL}}%
\setlength{\labelsep}{2. em}%
\setlength{\leftmargin}{\labelwidth}%
\addtolength{\leftmargin}{\labelsep}%
\setlength{\rightmargin}{0. cm}%
\setlength{\parsep}{\parskip}%
\setlength{\itemsep}{0. cm}\setlength{\topsep}{0. cm}%
\setlength{\partopsep}{0. cm}}

\Item{$f(\theta)$}{shape factor}
\Item{$\Delta g$}{nucleation driving force}
\Item{$k_{B}$}{Boltzmann's constant}
\Item{$k_{H}$}{Henry's constant}
\Item{$p$}{pressure}
\Item{$r$}{radius, radial coordinate}
\Item{$m_0$}{mass of one atom or molecule}
\Item{$n$}{number of atoms in a bubble}
\Item{$t$}{time}
\Item{$\textbf{v}$}{velocity}
\Item{$x$}{atomic fraction}
\Item{$C$}{concentration}
\Item{$D$}{diffusivity}
\Item{$G$}{Gibbs free energy}
\Item{$J$}{diffusion rate, depletion rate}
\Item{$M$}{molar mass}
\Item{$N_{A}$}{Avogadro's number}
\Item{$N_b$}{number of bubbles per unit volume}
\Item{$R$}{gas constant}
\Item{$S$}{source term, nucleation rate}
\Item{$T$}{temperature}
\Item{$Z$}{compressibility factor}
\end{list}

\subsection*{Subscripts}
\begin{list}{}{%
\settowidth{\labelwidth}{\textbf{HCLL}}%
\setlength{\labelsep}{2. em}%
\setlength{\leftmargin}{\labelwidth}%
\addtolength{\leftmargin}{\labelsep}%
\setlength{\rightmargin}{0. cm}%
\setlength{\parsep}{\parskip}%
\setlength{\itemsep}{0. cm}\setlength{\topsep}{0. cm}%
\setlength{\partopsep}{0. cm}}

\Item{$b$}{bubble}
\Item{$c$}{cluster}
\Item{$G$}{gas phase}
\Item{$He$}{helium}
\Item{$HEN$}{heterogeneous nucleation}
\Item{$HON$}{homogeneous nucleation}
\Item{$L$}{liquid bulk phase}
\Item{$nuc$}{nucleation}
\Item{$PbLi$}{lithium lead \LiPb\ eutectic}
\Item{$sur$}{surface}
\Item{$th$}{thermal}
\Item{$vol$}{volume, per unit volume}
\end{list}

\subsection*{Superscripts}
\begin{list}{}{%
\settowidth{\labelwidth}{\textbf{HCLL}}%
\setlength{\labelsep}{2. em}%
\setlength{\leftmargin}{\labelwidth}%
\addtolength{\leftmargin}{\labelsep}%
\setlength{\rightmargin}{0. cm}%
\setlength{\parsep}{\parskip}%
\setlength{\itemsep}{0. cm}\setlength{\topsep}{0. cm}%
\setlength{\partopsep}{0. cm}}

\Item{$m$}{molar}
\Item{$sat$}{saturation}
\Item{$xp$}{atomic fraction - pressure}
\Item{$0$}{pre-exponential}
\Item{$*$}{critical}
\end{list}

%%%%%%%%%%%%%%%%%%%%%%%%%%%%%%%%%%%%%%%%%%%%%%%%%%%%%%%%%%%%%%%%%%
%%% INTRO
%%%%%%%%%%%%%%%%%%%%%%%%%%%%%%%%%%%%%%%%%%%%%%%%%%%%%%%%%%%%%%%%%

\section{\label{sec:intro}Introduction}

Future magnetic confinement D--T fusion reactors, based on liquid metal (LM) eutectic alloy \LiPb as a coolant and breeding material, are supposed to be tritium fuel self-sufficient. Tritium production inside the so called breeding blankets is linked to He production in the LM, that may lead, under the necessary conditions, to nucleation events \citep{Batet}. Nucleated He bubbles may have a large impact on the self-sufficient principle, heat exchange, T permeation (leakage) and auxiliary systems. The present work intend to be a step forward toward the understanding of the complex phenomena that take place in a breeding blanket, focusing on He nucleation.

Homogeneous nucleation (HON), that is bubble formation in the bulk fluid, turns out to be triggered by neutron irradiation (radiation induced displacements in the LM structure), fluid discontinuities or temperature local peaks. Evidence of such phenomenon has been exposed in \citet{Conrad}, where impact on LM properties and T breeding ratio has been experimentally assessed. A thermodynamically self--consistent nucleation model including radiation effects has not been developed yet. However, many efforts towards the developments of such model have been made for solid irradiated materials, e.g., \citet{Trinkaus} and references there. Nucleation in LM under neutron irradiation studies are scarce as well as experimental data. Molecular dynamics (MD) studies on He cavitation in liquid lead for \LiPb phenomenon determination have been carried out by \citet{Bazhirov}: results show significant discrepancies with respect to CNT, which is not acceptable as it underestimates the work of formation due to the surface tension approximation to that of a planar surface. \citet{Bazhirov} state that the surface tension of a critical cluster, that is a stable cluster that will develop into a gas bubble, has a larger surface tension than that of the planar surface. This fact fully agrees with \citet{Tolman} surface tension correction, which predicts that surface tension for a droplet increases for increasing droplet sizes, while surface tension of a bubble decreases for increasing bubble sizes.

As has already been mentioned, in a Fusion reactor T is expected to be generated in order to fulfil fuel self-sufficiency requirements at the same rate than He (of the order of 500~g/day for a 3~GW$_{\textrm{th}}$ DEMO reactor \citep{Nishi}) due to the following nuclear reactions:
\begin{equation}
\mathrm{^{6}Li+n\rightarrow {^{3}H}+{^4He}+4.78\;MeV}\label{eq:reactLi6}.
\end{equation}
\begin{equation}
\mathrm{^{7}Li+n\rightarrow {^{3}H}+{^4He}+n-2.47\;MeV}\label{eq:reactLi7}.
\end{equation}

Anticipated results \citep{Batet}, using classical Nucleation Theory (CNT) showed that He nucleation event, rather than being a remote possibility, may occur under nominal conditions leading to a significant set of effects: flow regime perturbations, heat transfer efficiency reduction, degradation of pumping systems and T permeation reduction. Hence, the issue of He bubbles formation may be highly relevant to fusion reactor design and operation. He bubbles show up to act as a T sink, that may have an impact on T inventory as well as in T breeding ratio. Note that He bubbles may also have a large impact on T effective solubility, as T would be absorbed into the bubbles allowing more T to be present in the bulk LM.

The main aim of this paper is to give insight and to provide a reliable computational tool to quantify the He complex phenomena in a HCLL, in order to assess its potential effect. In the present work, a model based on the self--consistent nucleation theory (SCT) by \citet{Girshick90} is exposed, together with some other major improvements regarding radiation induced nucleation modelling and surface tension corrections. Implementation of the model have been done in the open source CFD code \OF (see \citet{Jasak} and references there) solver.

Note that the presented results deal with nano to micro bubbles upon their formation; at onset conditions and immediately after. Hence, the effect of the bubbles on the LM properties is out of the scope of the presented work and may deserve a dedicated publications taking into account multi phase flow. However it is worth to be noted that if bubbles become large enough, which is not a remote possibility under fusion conditions, LM effective density and viscosity may be affected by a dispersed He gas phase.

%%%%%%%%%%%%%%%%%%%%%%%%%%%%%%%%%%%%%%%%%%%%%%%%%%%%%%%%%%%%%%%%%
%%% MODEL
%%%%%%%%%%%%%%%%%%%%%%%%%%%%%%%%%%%%%%%%%%%%%%%%%%%%%%%%%%%%%%%%%

\section{H\lowercase{e} Nucleation Model}\label{sec:HeModel}

Simple models for He nucleation, bubble growth and transport, together with T complex transport phenomena, have been previously developed and implemented in the \OF CFD code \citet{Batet}.

Homogeneous nucleation mechanism can be summarized as follows:

\begin{itemize}
\setlength{\itemsep}{1pt}
\setlength{\parskip}{0pt}
\setlength{\parsep}{0pt}
 \item{Helium concentration in the LM bulk increases due to nuclear reactions Eq.~\ref{eq:reactLi6} and Eq.~\ref{eq:reactLi7}.}
 \item{Non-stable He clusters begin to form due to high He concentration in the bulk.}\\
 \item{Non-stable He clusters become larger as He concentration in the LM bulk increases.}\\
 \item{Eventually, non-stable clusters overcome the nucleation barrier and He clusters become stable.}\\
 \item{Clusters grow until a new phase arises in the form of bubbles.}\\
\end{itemize}

In the present section some major improvements to the homogeneous nucleation (HON) model are exposed.

\subsection{SCT Homogeneous Nucleation Model}\label{subsec:HON}

The CNT HON rate is expressed as follows (see \citet{Batet} and references therein):
\begin{equation}
S_{CNT,HON}\ =\ S^{0}_{HON}\,e^{-\Delta G^{\ast}_{HON}/k_{B}T}\label{eq:CNTHON}
\end{equation}
where $S^{0}_{HON}$ is a pre-exponential factor that depends on the kinetics of the system, $\Delta G^{\ast}_{HON}$ is the barrier height to nucleation, $k_{B}$ is the Boltzmann constant and $T$ is the liquid metal bulk temperature. An expression for the pre-exponential factor, deduced by \citet{Oxtoby}, is used:
\begin{equation}
S^{0}_{HON}\ =\ \frac{\upsilon_0\, N_A^{\, 2}\,C_{He,LM}^{\, 2}}{\psi}\, \biggl(\frac{2\sigma}{\pi m_{0}}\biggl)^{1/2}\label{eq:Preexp}
\end{equation}
where $N_A$ is the Avogadro's number, $C_{He,LM}$ the He concentration in LM, $\psi$ is the supersaturation ratio, relating the actual He concentration with the saturation concentration, $\upsilon_0$ and $m_{0}$ are the volume and the mass of one He atom in the cluster, respectively. It should be mentioned that similar analytical expressions for $S^{0}_{HON}$ can be found in \citet{Wu} and \citet{Kwak}.

CNT was modified by \citet{Katz} into the kinetic nucleation theory (KNT) and extended by \citet{Girshick90} to derive a new expression for the HON rate referred to the stable equilibrium of a supersaturated vapor; the work of formation of one monomer is assumed to be zero. The work of formation $\Delta G$  of a cluster of radius $r_{c}$ for CNT and SCT are the following:
\begin{equation}\label{eq:DGtotCNT}
\Delta G_{CNT}\ =\ \Delta G_{sur}\ +\ \Delta G_{vol}\ =\ 4\pi r_{c}^2\sigma\ +\ \dfrac{4}{3}\pi r_{c}^3\Delta g_{vol}
\end{equation}
\begin{equation}\label{eq:DGtotSCT}
\Delta G_{SCT}\ =(\ 4\pi r_{c}^2-s_0)\sigma\ +\left[ \left( \dfrac{4}{3}\pi r_{c}^3\right)-1\right]\Delta g_{vol}
\end{equation}
where $\Delta G_{sur}$ and $\Delta G_{vol}$ are the gain in free energy of the new stable phase and the cost in free energy due to the introduction of the interface, $\Delta g_{vol}$ is the driving force for nucleation per unit volume of the new phase and $s_0$ is the surface area of one He atom in the cluster.

The remaining physical assumptions in SCT are identical to those used in the CNT. \citet{Girshick91} derived the same rate directly from the CNT, resulting in a rather simple modification of the CNT HON nucleation rate (Eq.~\ref{eq:CNTHON}):
\begin{equation}
S_{SCT,HON}\ =\ \dfrac{e^\Theta}{\psi}\, S_{CNT,HON}\label{eq:SCTHON}
\end{equation}
being $\Theta$ the surface energy of one He atom in the cluster:
\begin{equation}
\Theta\ \equiv\ \dfrac{\sigma s_0}{k_{B}T}\label{eq:SCTcorr}
\end{equation}

Note that SCT rate correction (Eq.~\ref{eq:SCTcorr}) is highly dependent on $T$ and strongly sensitive to $s_0$.

SCT has gained acceptance due to its good results, reliability and simplicity \citep{Girshick90}.

\subsection{Irradiation induced nucleation model}\label{subsec:sHON}

Neutron radiation lowers the barrier height to nucleation, which is the necessary condition to nucleation, due to the local energy deposition, radiation damage or displacements and He local accumulation due to nuclear reactions (Eq.~\ref{eq:reactLi6} and Eq.~\ref{eq:reactLi7}). In the present section a novel simple model is presented so as to take into account this phenomenon.

The barrier height to nucleation reads,
\begin{equation} \label{eq:DGcrit}
\Delta G^{\ast}_{HON}\ =\ \frac{16 \pi \sigma^3}{3\ \Delta g_{vol}^2}
\end{equation}
where $\Delta g_{vol}$ is expressed by \citep{Wu,Frenkel,Gunton}: 
\begin{equation} \label{eq:Dgvol}
\Delta g_{vol} = \frac{- k_B T}{\upsilon_0} \ln(\psi)
\end{equation}
Radiation barrier lowering might be taken into account through Heterogeneous nucleation (HEN) (\citep{Batet}, that is nucleation on a substrate like surfaces or impurities, as this phenomenon lowers the nucleation barrier. Bubbles are treated as spherical caps nucleating in the bulk fluid, which is a very rough estimation for the bubble area and volume, with a significant impact on growth rates. In the present work, barrier height reduction is taken into account through a parameter $b$ that has to be determined experimentally, resulting into spherical bubbles and more accurate bubble surface areas. Moreover, present model represents the physical phenomenon with much more accuracy. Note that $b$, fit to experimental data, will include not only radiation effect but all nucleation triggering phenomena with a single parameter. Under fusion conditions, nuclear irradiation is not to be constant over time, thus, $b$ parameter will be not only a function over time, but also a function of the nuclear depositions (i.e. function of the He concentration source term in the bulk LM).

The following expression for the barrier height to nucleation is proposed:
\begin{equation} \label{eq:DGcritcorr}
\Delta G^{\ast}_{HON}\ =\ b \frac{16 \pi \sigma^3}{3\ \Delta g_{vol}^2}
\end{equation}
Note that for $b$=0.5 the same barrier height reduction as for HEN assuming spherical cap bubbles is obtained ($\Delta G^{\ast}_{HON}>\Delta G^{\ast}_{HEN}$). Despite the same barrier height reduction, nucleation rates will be very different as thr pre-exponential factors for HON and HEN are different.

\subsection{Surface tension curvature correction models}\label{subsec:surfTen}

One of the major drawbacks of the CNT is that the surface tension of a nucleating bubble is approximated to that of a plane surface, resulting in an unacceptable underestimation of the nucleation rate. Barrier height to nucleation is a strong function of the surface tension (see. Eq.~\ref{eq:DGcritcorr}), which means that taking the surface tension of a growing bubble as constant may not be acceptable for small bubble radii (under the micrometric scale). A small uncertainty in the surface tension may slightly change the surface energy, but will dramatically change the nucleation rate as it is exponentially dependent on the surface energy. \citet{Tolman} derived a simplified relation between the surface tension and the radius of growing bubbles:
\begin{equation}
\sigma=\dfrac{\sigma_0}{1-\dfrac{2\delta_T}{r_{b}}}\label{eq:Tolman}
\end{equation}
where $\delta_T$ is the so called Tolman's length, which is a constant defined as the distance between the bubble surface of tension and the equimolar surface inside the interfacial region $\delta_T=r_{eq}-r_{b}$ and is commonly set between 0.25 and 0.5 of the diameter of the gas phase species. Despite that there is neither a thermodynamically consistent theoretical model nor experimental data on the value of $\delta_T$, some approximations can be found in the literature (see,e.g., \citet{Blokhuis}, \citet{Moody_03}).

In the present work the following model is implemented:
\begin{equation}
\delta_T^\prime=2\ a\ r_0 =a\ \delta_T\label{eq:delta}
\end{equation}
where $a$ is an adjustable parameter ($a<0$, see e.g., \citet{Tolman} and \citet{Moody_03}) and $r_0$ is the atomic radius of the gas species in a cluster.

Note that some authors have presented $\delta_T$ as a function of the radius (\citep{Blokhuis}, \citep{Moody_01}, \citep{Bartell} and \citep{Vogelsberger} among others). In the present work, Tolman length has been taken as a constant in order to have a simple and suitable model for CFD implementation: $\delta_T$ functions involve iterative processes and complex calculations that will dramatically increase computational costs as well as introduce numerical instabilities.

\citet{Rasmussen} proposed the following expression for the relation between the surface tension and the radius of a growing bubble:
\begin{equation}
\sigma=\sigma_0\left( 1+\dfrac{\delta_T}{r_{b}} \right) ^2\label{eq:Rasmussen}
\end{equation}

In the present work a generalized Tolman expression with two adjustable parameters $a$ and $c$ is proposed:
\begin{equation}
\sigma=\sigma_0\left( 1+\dfrac{a\delta_T}{r_{b}} \right) ^c=\sigma_0\left( 1+\dfrac{\delta_T^\prime}{r_{b}} \right) ^c\label{eq:General}
\end{equation}

It should be noted that as a bubble becomes smaller, the surface tension increases, and so it does the barrier height to nucleation. As has already been pointed out, \citet{Bazhirov} MD results predict greater surface tensions for nucleating bubbles than those for the planar surface (negative Tolman's length) resulting into lower nucleation rates.

\subsection{Kinetic--Diffusion growth model}\label{subsec:growth}

Bubble growth can be understand as a pure diffusion growth model, assuming that once a clusters reaches its critical radius it becomes a bubble with macroscopic gas phase properties. This assumption holds only if a critical cluster grows fast enough to reach the necessary radius to become a macroscopic bubble. However, the transition from the critical cluster to a bubble big enough to be considered a macroscopic gas bubble, hereinafter called critical bubble, might not be sufficiently fast to be neglected.

\begin{figure}[h!]
\begin{center}
\includegraphics[width=1\columnwidth]{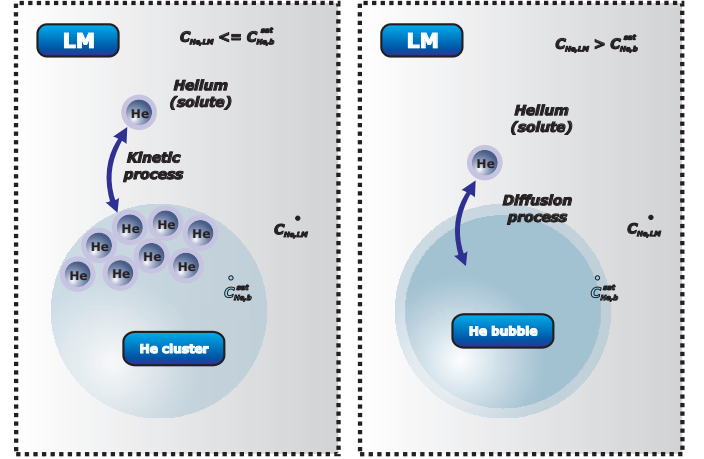}
\caption{Kinetic growth model (left) and diffusion growth model (right) showing when a model is used depending on the He concentrations.}
\vspace{-10pt}
\label{fig:phenomena}
\end{center}
\end{figure}

Bubbles will shrink to disappear if a diffusion--controlled growth model is used for critical clusters, due to the fact that inner cluster pressure is so high with respect to LM pressure ($p_b \gg p_{LM}$) that diffusion process takes place towards the LM, that is removing He from the cluster. Diffusion process will only take place towards the bubble (Fig.~\ref{fig:phenomena} left)if the relation between bulk concentrations is $C_{He,ib} < C_{He,LM}$, assuming that $C_{He,ib} = C_{He,b}$ and $C_{He,iLM} = C_{He,LM}$, where $i$ denotes the interface.

Assuming mechanical and chemical equilibrium between phases, and following the approach exposed in \citet{Kwak}, the simplification of the \citet{Epstein} model (diffusion process)  applies only if $C_{He,b}^{sat} < C_{He,LM}$, where $C_{He,b}^{sat}$ is referred to the bubble's pressure and volume according to the Henry's law.

The transition from the critical cluster to the critical bubble can be modelled as a kinetic controlled growth (Fig.~\ref{fig:phenomena} right). A critical cluster will grow each time a He atom strikes the cluster surface and stays stuck. The He striking rate, in mol/s, can be expressed through the kinetic theory of gases as follows (see \citet{Kwak}, \citet{Swandic} and \citet{Noyes} for more details on the kinetic model derivation):
\begin{equation}
J_{He,b}^{KT}=\dfrac{\beta}{4}\left(\dfrac{8\ k_B\ T}{\pi\ m_0}\right)^{1/2} = \dfrac{\beta p_{LM}}{(2\ \pi\ m_0\ k_B\ T)^{1/2}}\label{eq:growthKT}
\end{equation}
where $\beta$ is the accommodation parameter, that is the probability of an atom to stay stuck after a collision with a cluster.

In the present work both, diffusion and kinetic models are used as follows:
\begin{equation}
J_{He,b}= \left\{ 
\begin{matrix}
J_{He,b}^{KT} = \dfrac{\beta p_{LM}}{(2\ \pi\ m_0\ k_B\ T)^{1/2}}&\,& C_{He,b}^{sat} \geq C_{He,LM}\\[5mm]
J_{He,b}^{Diff} = 4\pi r^{2}_{b}\, D_{He,LM}\,\biggl(\dfrac{\partial C_{He,LM}}{\partial r}\biggl)_{r=r_{b}}&\,&C_{He,b}^{sat} < C_{He,LM}\\[5mm]
\end{matrix}\right.
\end{equation}
where $D_{He,L}$ is the diffusion coefficient. The concentration gradient $\left( \partial C_{He,LM}/\partial r \right)_{r=r_{b}}$ is approximated to: 
\begin{equation}
\biggl(\dfrac{\partial C_{He,LM}}{\partial r}\biggl)_{r=r_{b}}\approx \frac{C_{He,L}-C^{sat}_{He,b}}{r_{b}}\label{eq:Cgrad}
\end{equation}

\section{Results and Discussion}

Model improvements have been implemented in the OpenFOAM\textsuperscript{\textregistered} toolbox, which uses the finite volume method, as a part of the solver. The new solver is applied as a post-process to the hydrodynamics solution assuming there is no effect of the bubbles on the LM velocity field and properties. Bubble size is calculated using the mean radius approach (MRA), averaging the size of new-born bubbles with those already present in the LM.

In the following subsections 
Case of interest for Fusion technology
Comparison between models
Hint or guess about possible scenarios that raise technical implications

The model has been applied to a zero-dimensional domain, in order to evaluate its performance as an example of code capabilities, and compared to former model results.

\subsection{Zero Dimensional Analysis}\label{zeroD}

A single cell containing \LiPb\ is simulated. Velocities are set to zero (no convection). Atomic helium generation rate is set to a constant value of 10$^{-7}$~mol/(m$^3$s); pressure and temperature are set to constant values of 2~bar and 723.15~K. Diffusivity and solubility, together with other material properties, have been taken from the \LiPb database for nuclear fusion technology (\citet{MasdelesValls}) at HCLL breeding blanket operation conditions:

\begin{list}{}{%
 \setlength{\labelwidth}{0.3 \linewidth}%
 \setlength{\labelsep}{1. ex}%
 \setlength{\leftmargin}{\labelwidth}%
 \addtolength{\leftmargin}{\labelsep}%
 \setlength{\rightmargin}{0. cm}%
 \setlength{\parsep}{\parskip}%
 \setlength{\itemsep}{0. cm}\setlength{\topsep}{1.ex plus0.2ex minus0.1ex}%
 \setlength{\partopsep}{0. cm}}
   \Itema{$M_{PbLi}=$}{0.17316 kg/mol}
   \Itema{$\rho_{PbLi}=$}{9660 kg/m$^3$}
   \Itema{$\sigma_{0,He,PbLi}=$}{0.46 kg/s$^2$}
   \Itema{$k_{H}^{xp}=$}{3 10$^{-14}$ mol$_{He}$/(mol$_{PbLi}$Pa)}
   \Itema{$M_{He}=$}{0.004 kg/mol}
   \Itema{$D_{He,PbLi}=$}{5 10$^{-8}$ m$^2$/s}
\end{list}

Concerning the helium atomic volume, $\upsilon_0$, the value for He clusters in \LiPb can not be found in literature. As a base case calculation, He atomic volume is set to 1.7~10$^{-29}$~m$^3$, within the range of empirical values for solid metals (see, e.g., \citet{Donnelly}). Surface area of He atoms in a cluster $s_0$ is directly calculated from $\upsilon_0$ with a value of 3.1972~10$^{-19}$~m$^2$. Onset rate threshold has been arbitrarily chosen at 1~m$^{-3}$s$^{-1}$ (setting an arbitrary threshold over which nucleation rate is considered significant is needed by the nature of equation~\ref{eq:CNTHON} and~\ref{eq:SCTHON}; e.g., \citet{Goldman} uses 1~cm$^{-3}$s$^{-1}$). For HEN, a pre-exponential factor of 10$^{21}$~bubbles/(m$^3$s) and a contact angle between the nucleation substrate and the bubbles $\theta=\pi/2$ (spherical cap bubbles) is set.

Parameter $b$ is arbitrarily set to 0.5, which means that nucleation triggering phenomena reduces the barrier height to one half that of pure HON (constant $b$=1), in order to compare HEN in \citep{Batet} to present model results for CNT and SCT. Note that no surface tension correction has been used in this base case simulation ($\sigma=\sigma_0$). 
\begin{figure}[h!]
\begin{center}
\includegraphics[width=1\columnwidth]{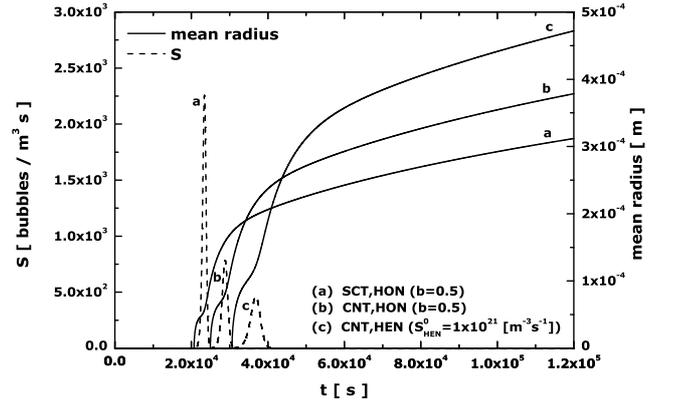}
\caption{Nucleation rate (dashed lines, left) and mean radius (solid lines, right) comparison between CNT (\textbf{a} HON, \textbf{b} HEN) and SCT (\textbf{c}) models.}
\vspace{-10pt}
\label{fig:SCTSens1}
\end{center}
\end{figure}
\begin{figure}[h!]
\begin{center}
\includegraphics[width=1\columnwidth]{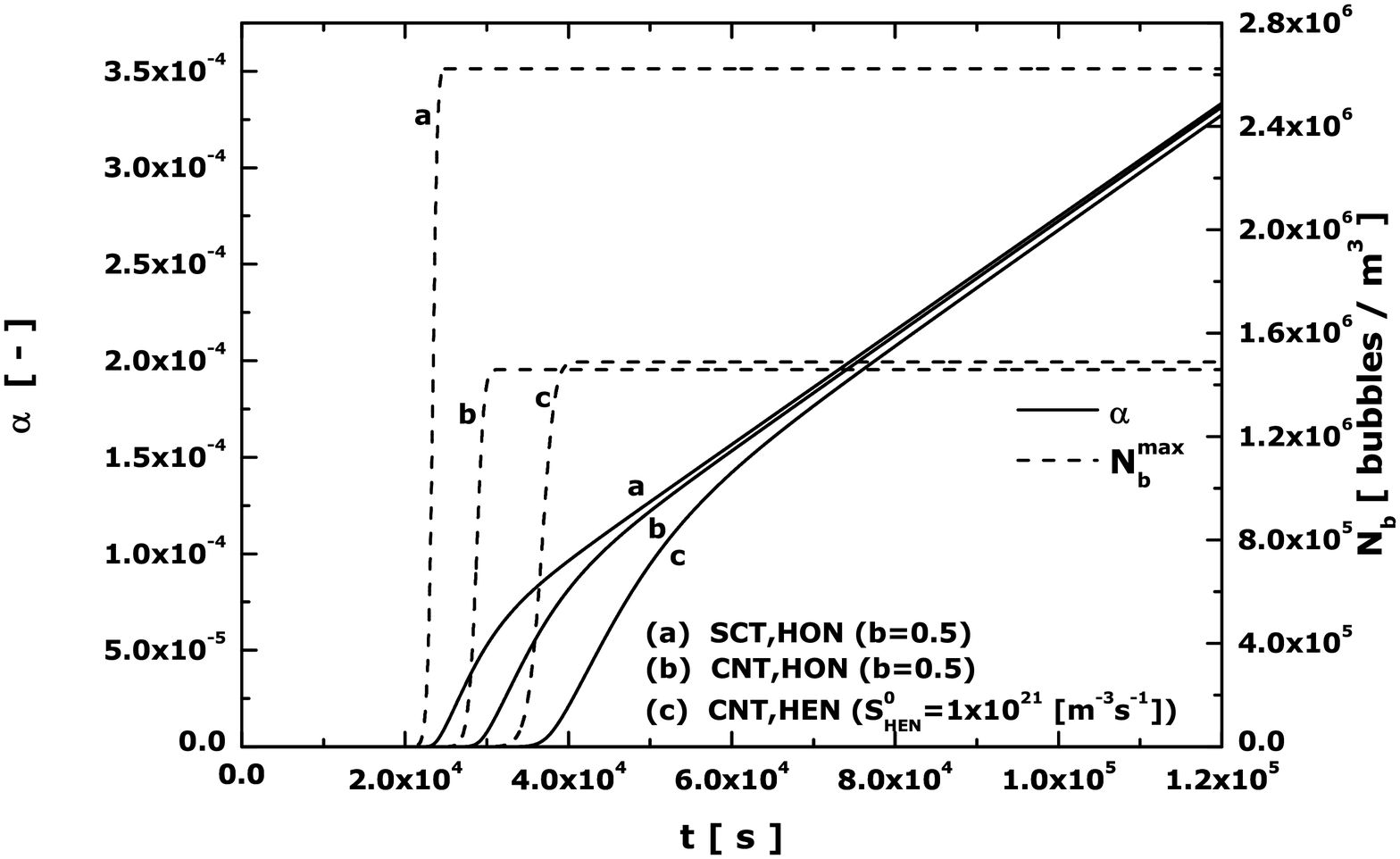}
\caption{Void fraction $\alpha$ (dashed lines, left) and bubble concentration (solid lines, right) comparison between CNT (\textbf{a} HON, \textbf{b} HEN) and SCT (\textbf{c}) models.}
\vspace{-10pt}
\label{fig:SCTSens2}
\end{center}
\end{figure}

SCT predicts earlier nucleation onset time than CNT (see Fig.~\ref{fig:SCTSens1} and Fig.~\ref{fig:SCTSens2}), with higher rates as expected and already noted in \citep{Girshick90}. This later fact implies that nucleation under fusion reactor conditions may happen even sooner than expected and reinforces the necessity to conduct experiments for this type of systems. Although HEN and HON with constant $b=$~0.5 over time have the same barrier height, pre-exponential factors are different, so HEN nucleates the latest.

Radius is significantly affected depending on which model, CNT or SCT, is used. SCT onset occurs much earlier than for CNT, so more bubbles are formed reaching smaller radius. CNT, though, predicts larger onset times. Thus, nucleating bubbles find a large concentration of dissolved He which results in less bubbles and higher growth rates. 
\begin{figure}[h!]
\begin{center}
\includegraphics[width=1\columnwidth]{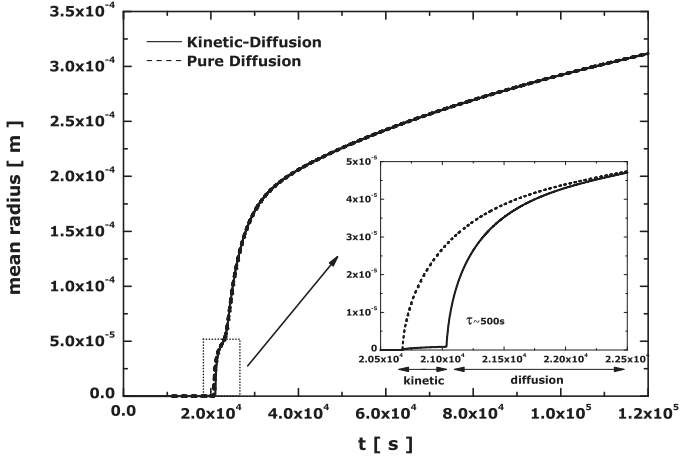}
\caption{SCT pure diffusion (dashed line) and kinetic--diffusion (solid line) models mean radius comparison for the nucleation event.}
\vspace{-10pt}
\label{fig:GrowthModels}
\end{center}
\end{figure}

For the SCT model, Fig.~\ref{fig:GrowthModels} shows the radius evolution comparison between a pure diffusion growth model and the present work model for $\beta=$1. Major differences are shown right after nucleation onset: growth due to kinetic model is much slower than that of the pure diffusive model. After a short time the critical bubble condition is met and bubbles begin to grow due to diffusion. The longer the bubbles stay growing by diffusion the smaller the mean radius difference between both models becomes. The kinetic--diffusion model delays the bubble growth with respect to the pure diffusion model. However, for long simulation times this effect turns out to be negligible, as both models tend towards the same value asymptotically.

Some authors, e.g, \citet{Kashchiev} and \citet{Shneidman}, introduce the idea of a lag time $\tau$ (incubation time, delay time, etc.), which is the mean time critical clusters need to reach the critical bubble size. For the SCT kinetic-diffusion growth model simulation in Fig.~\ref{fig:GrowthModels} a $\tau\approx500s$ is found. Lag times strongly depend on system's properties and the supersaturation ratio; \citet{Lamer} found lag times in the range of $0.01<\tau<3000$ for barium sulfate precipitation at different liquid bulk concentrations while \citet{Kwak85} found lag times of a fraction of a second for water--gas systems.

\subsection{SCT Sensitivity Analyses}

A series of sensitivity analyses to key system parameters have been performed by means of zero-dimensional calculations.

\subsubsection{Barrier Height to Nucleation Parameter $b$}

\begin{figure}[h!]
\begin{center}
\includegraphics[width=1\columnwidth]{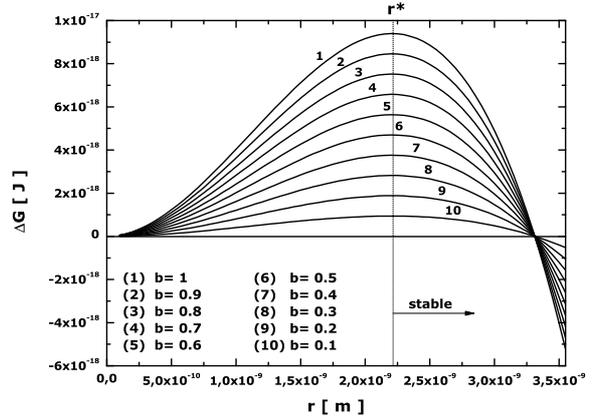}
\caption{Nucleation barrier height $\Delta G^{\ast}$ ($r=r^{\ast}$)sensitivity to irradiation induced parameter $b$. Bubble is assumed to be stable for $r>r^{\ast}$}
\vspace{-10pt}
\label{fig:DGbSens}
\end{center}
\end{figure}

\begin{figure}[h!]
\begin{center}
\includegraphics[width=1\columnwidth]{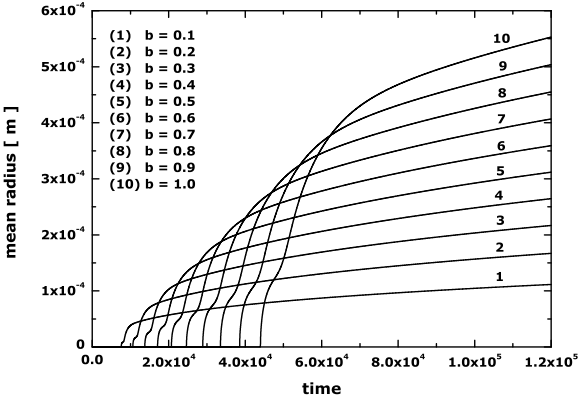}
\caption{Mean bubble radius sensitivity to irradiation induced parameter $b$ for the nucleation event.}
\vspace{-10pt}
\label{fig:DGbSensRm}
\end{center}
\end{figure}

\begin{figure}[h!]
\begin{center}
\includegraphics[width=1\columnwidth]{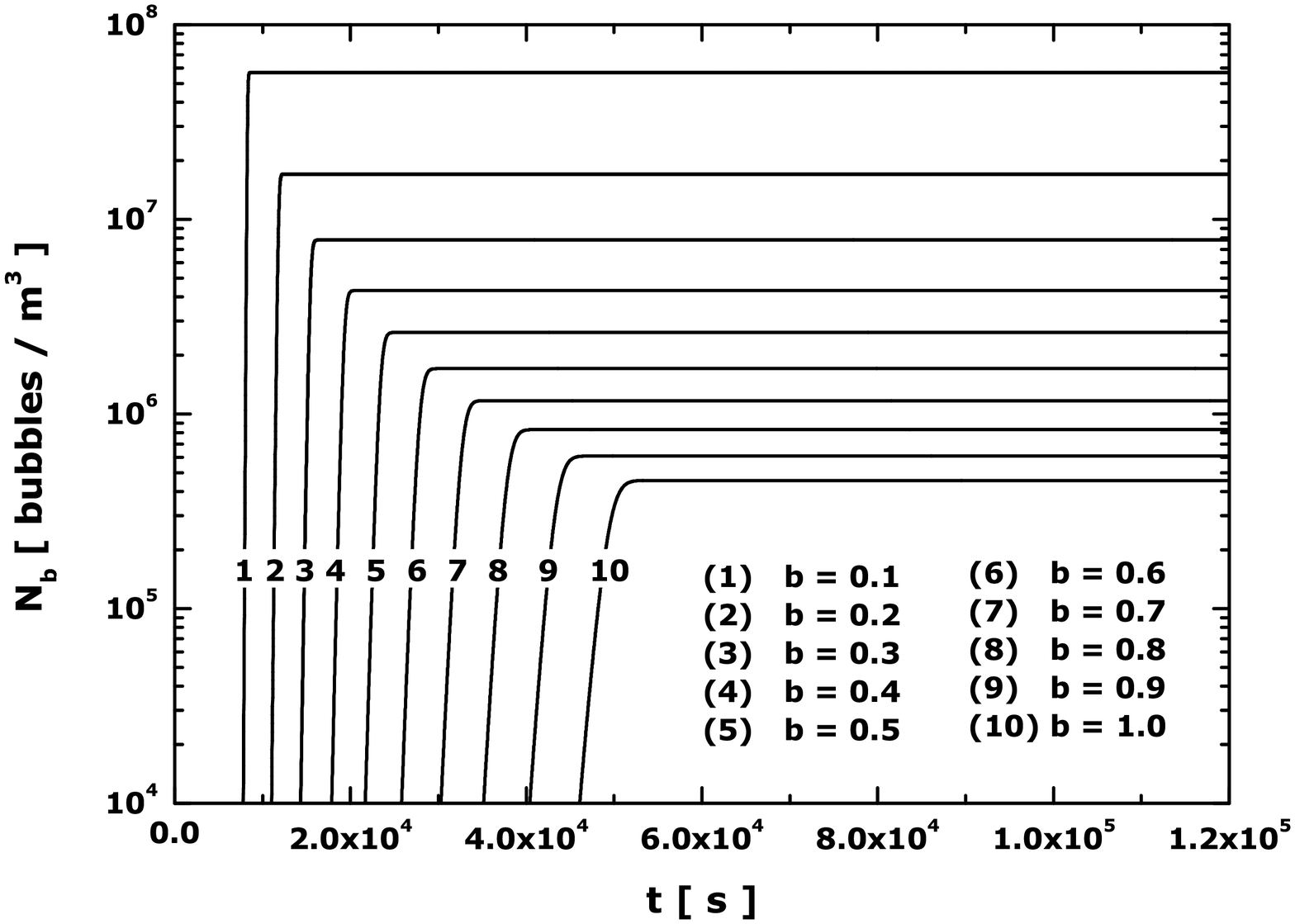}
\caption{Bubble concentration sensitivity to irradiation induced parameter $b$ for the nucleation event}
\vspace{-10pt}
\label{fig:DGbSensNb}
\end{center}
\end{figure}

\begin{figure}[h!]
\begin{center}
\includegraphics[width=1\columnwidth]{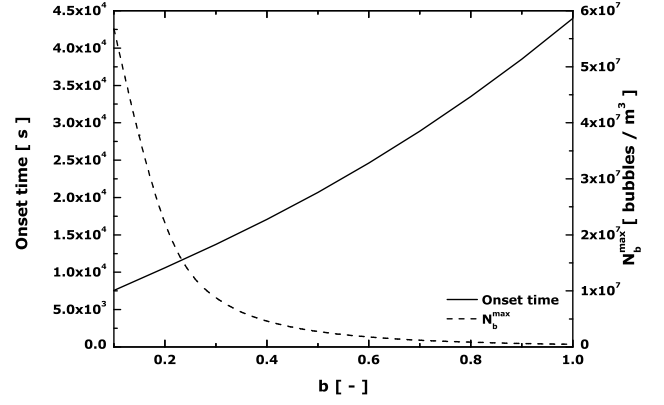}
\caption{Onset time (solid line, left) and maximum concentration of bubbles (dashed line, right) sensitivity to irradiation induced parameter $b$.}
\vspace{-10pt}
\label{fig:SCTbSens}
\end{center}
\end{figure}

For the same base case conditions in Sec.~\ref{zeroD}, Fig.~\ref{fig:DGbSens} shows how the barrier height to nucleation is reduced as a function of $b$. The lower the barrier height is the lower supersaturation $\psi$ is needed to have stable clusters. Therefore, less time is needed to reach onset time, but as there is less He concentration in the liquid, less bubbles are nucleated (see Fig.~\ref{fig:SCTbSens}). For high $b$ values a small number of bubbles nucleate in a rich dissolved He media. Hence, each bubble absorbs a large amount of He from the liquid, reaching a bigger radius than a bubble that nucleates at low $b$ values as shown in Fig.~\ref{fig:DGbSensRm}. For low $b$ values a huge amount of bubbles nucleate (see Fig.~\ref{fig:DGbSensNb}), which means that each bubble absorbs less He (lower growth rates). Lowering $b$ one order of magnitude results in a change of five orders of magnitude in the nucleation rate. $b$ parameter turns out to be an effective way to fit nucleation onset time to experimental data if available, as it fully controls the necessary condition to nucleation $\Delta G^{\ast}$. Note that $b$ has been kept constant over time for simplicity and to show the pure effect of this parameter. Under real fusion conditions, $b$ would be a function of time and He concentration source term, i.e. the amount of nuclear irradiation on the bulk liquid.


 Despite the fact that preliminary MD model and calculations, from \citet{Bazhirov}, agree with exposed results, parameter $b$ should be determined by means of complex MD simulations. An accurate He-PbLi interatomic potential describing the compound is needed \citep{Fraile}, being such development out of the scope of the present work.
For instance, it is worth noting that MD simulation for different amounts of nuclear irradiation will give some insight on the nucleation event behavior. Moreover, simulating fusion conditions through ramp irradiations, would give insight on the expected amount of bubbles in a LM blanket loop of a fusion reactor.

\subsubsection{System's Temperature}

\begin{figure}[h!]
\begin{center}
\includegraphics[width=1\columnwidth]{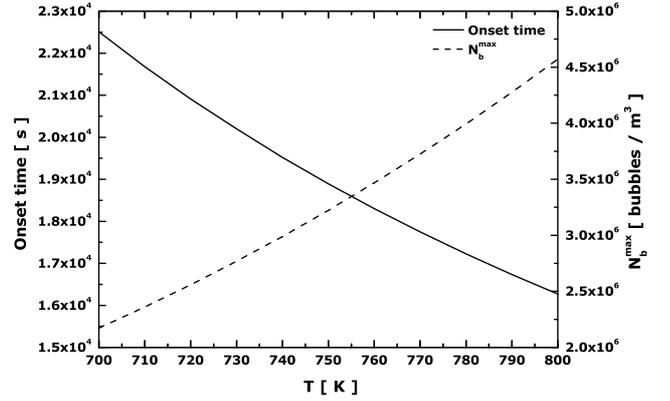}
\caption{Onset time (solid line, left) and maximum concentration of bubbles (dashed line, right).}
\vspace{-10pt}
\label{fig:tempSens}
\end{center}
\end{figure}

A sensitivity analysis to system's temperature has been performed around the LM breeder nominal conditions. Despite SCT model is strongly dependent on the temperature, results in Fig.~\ref{fig:tempSens} give a change of less than one order of magnitude in the onset time and the maximum number concentration of bubbles for a change of one order of magnitude for the temperature. Results, however, show that temperature dependency is not negligible.

\subsubsection{Curvature Corrections to the Surface Tension}

A sensitivity analysis to the surface tension curvature correction models is performed for the same base case conditions in Sec.~\ref{zeroD}.

\begin{figure}[h!]
\begin{center}
\includegraphics[width=1\columnwidth]{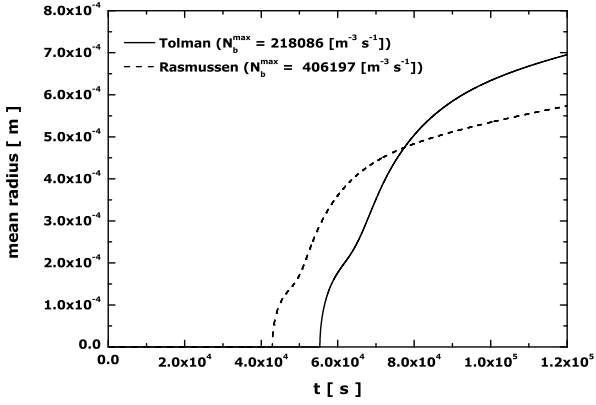}
\caption{Mean radius sensitivity to \citet{Tolman} (solid line) and \citet{Rasmussen}(dashed line) models for the nucleation event.}
\vspace{-10pt}
\label{fig:RmsSens}
\end{center}
\end{figure}

\begin{figure}[h!]
\begin{center}
\includegraphics[width=1\columnwidth]{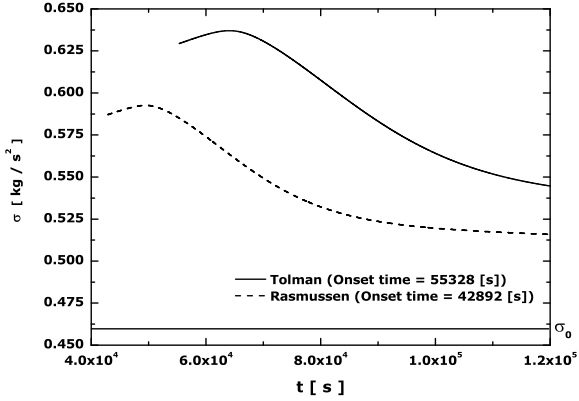}
\caption{bubble $\sigma$ sensitivity to \citet{Tolman} (solid line) and \citet{Rasmussen}(dashed line) models for the nucleation event.}
\vspace{-10pt}
\label{fig:sigmasSens}
\end{center}
\end{figure}

Differences between \citet{Tolman} and \citet{Rasmussen} models are significant: mean radius comparison in Fig.~\ref{fig:RmsSens} shows that Tolman's model for a constant $\delta_T=$~0.35 (see, e.g. \citep{Tolman} and \citep{Moody_01} for $\delta_T$ commonly used values) triggers nucleation later than for Rasmussen's model. Tolman's model predicts a higher surface tensions as shown in Fig.~\ref{fig:sigmasSens} leading to a higher barrier height. Hence, nucleation onset time is delayed.

After onset, critical radius of new--born bubbles decreases due to the fact that dissolved He is removed by absorption (bubble growth) and by bubble nucleation, which result in lower barrier heights. Therefore, surface tension of bubbles becomes larger as nucleation rate reaches its maximum. As has already been mentioned, nucleating bubble sizes are averaged with those already present in the LM, so as nucleation event reaches its end, due to dissolved He depletion, the weight of previously nucleated bubble sizes, when averaging, is larger than that of new ones. After nucleation event, surface tension grows asymptotically towards the surface tension of the planar surface $\sigma_0$.

\begin{figure}[h!]
\begin{center}
\includegraphics[width=1\columnwidth]{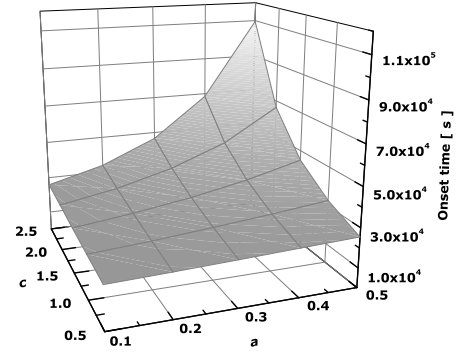}
\caption{Onset time sensitivity to present work's surface tension model parameters $a$ and $c$.}
\vspace{-10pt}
\label{fig:OnsetABSens}
\end{center}
\end{figure}

\begin{figure}[h!]
\begin{center}
\includegraphics[width=1\columnwidth]{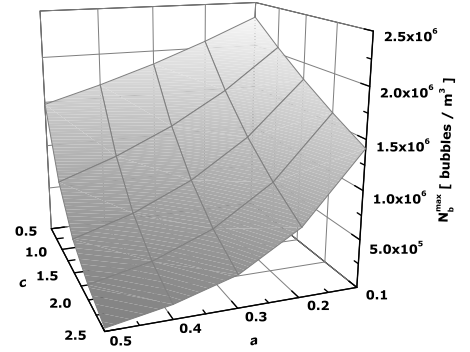}
\caption{Maximum concentration of bubbles sensitivity to present work's surface tension model parameters $a$ and $c$.}
\vspace{-10pt}
\label{fig:NbABSens}
\end{center}
\end{figure}

For Eq.~\ref{eq:General}, sensitivities to parameters $0.1 \leq a \leq 0.5$ and $0.5 \leq c \leq 2.5$ are shown in Fig.~\ref{fig:OnsetABSens} and Fig.~\ref{fig:NbABSens}. Parameter ranges have been set following already mentioned literature in Sec.~\ref{subsec:surfTen}.

%

%lbrt

As $a$ or $c$, or both, increase, nucleation begins later because surface tension becomes smaller; it should be kept in mind that $a$ directly modifies the Tolman length as exposed in Eq.~\ref{eq:General}. 

The sooner nucleation begins the more bubbles nucleate. This later tendency is clearly shown in Fig.~\ref{fig:NbABSens}, where maximum concentration of bubbles corresponds to $a=$~0.5 and $c=$~0.1. A change of one order of magnitude in either $a$ or $c$ leads to a change of two orders of magnitude in the maximum concentration of bubbles, which makes the models highly sensitive to any change in the He bubble-\LiPb surface tension. Eq.~\ref{eq:General} gives the necessary flexibility to fit any experimental data when available.

\section{Conclusions}

The work presented here exposes an improved model for HON in LM breeders of a fusion reactor (with respect to that exposed in \citep{Batet}), together with its implementation in the \OF CFD code solver. SCT has been successfully adapted for a CFD code with significantly different results that show onset times at lower supersaturation ratios. He bubbles mean radius is predicted to be smaller for HON with $b$=0.5 than for HEN, resulting in lower growth rates for HON. SCT model nucleation rate sensitivity to system's temperature around the operation conditions of a HCLL shows a significant dependence, which means that an isothermal approximation is not valid even if the system temperature varies in the range of tens of K around the nominal temperature.

Nucleation sensitivity to bubble surface tension as a function of the bubble radius shows a significant deviation with respect to the constant planar surface tension approximation used in our former model: onset times occur even at higher supersaturation ratios. However, it must be noted that SCT predicts earlier onset times, so SCT--$\sigma$ correction effects counteract each other to some extent.

A reference base case simulation is exposed in Fig.~\ref{fig:RmsSens} for SCT HON with $b=$~0.5 and Tolman's $\sigma$ correction model (particular case of Eq.~\ref{eq:General} with $c=$~1) with a constant $\delta_T=$~0.35. Result shows that nucleation event is delayed with respect to the CNT prediction. However, it still shows that nucleation phenomenon may take place under nominal conditions. This possibility may have a significant impact on tritium breeding ratio, that may be lower than expected. In addition, tritium inventory would be affected as tritium would be absorbed into Her bubbles.

Present work reinforces the necessity of conducting experiments to determine nucleation conditions and bubble transport parameters as well as the fact that nucleation may not be a remote phenomenon in LM breeders; experimental measures and MD simulations are necessary to improve and fully validate, which, despite its limitations, provides a good qualitative insight into the helium nucleation phenomenon in LM systems for fusion technology and can be used to identify key system parameters. Experiments to determine the nucleation conditions, parameter $b$, Tolman's $\sigma$ and growth models, are not necessary to be conducted under nuclear fusion conditions, as many of the models do not need those conditions to be determined. However, it is worth noting that experiment under nuclear fusion conditions would give a valuable insight on the whole phenomena.



%% References


\bibliographystyle{model1-num-names}
%%\bibliography{<your-bib-database>}





% \end{thebibliography}

\end{document}